\documentclass{aa}
\usepackage[T1]{fontenc}
\usepackage{amsmath}
\usepackage{txfonts}
\usepackage{amsmath}
\usepackage{graphicx}
\usepackage[active]{srcltx}
\usepackage{threeparttable}
\usepackage{natbib}
\usepackage{enumerate}
\bibpunct{(}{)}{;}{a}{}{,}

\newcommand{\niia}{[N\,\textsc{ii}]6584}

\newcommand{\oiii}{[O\,\textsc{iii}]5007}
\newcommand{\oii}{[O\,\textsc{ii}]3727}

\newcommand{\ha}{H$\alpha$}
\newcommand{\msun}{M$_{\odot}$}

\title{Integral field spectroscopy with SINFONI of VVDS galaxies\thanks{Based on observations collected at the European Southern Observatory (ESO) Very Large Telescope, Paranal, Chile, as part of the Programs 75.A-0318, 78.A-0177, and 070.A-9007}}
\subtitle{II. The mass-metallicity relation at $1.2 < z < 1.6$}

\author{J. Queyrel\inst{1}\and 
  T. Contini\inst{1}\and 
  E. P\'{e}rez-Montero\inst{1, 3}\and 
  B. Garilli\inst{4}\and 
  O. Le F\`evre\inst{2}\and 
  M. Kissler-Patig\inst{6}\and 
  B. Epinat\inst{1} \and 
  D. Vergani\inst{5} \and 
  L. Tresse\inst{2} \and 
  P. Amram \inst{2} \and
  M. Lemoine-Busserolle\inst{7}
}

\institute{
Laboratoire d'Astrophysique de Toulouse-Tarbes, Universit\'e de Toulouse, CNRS, 14 Avenue \'Edouard Belin, F-31400 Toulouse, France
\and 
Laboratoire d'Astrophysique de Marseille, Universit\'e de Provence, CNRS, 38 rue Fr\'ed\'eric Joliot-Curie, F-13388 Marseille Cedex 13, France
\and
Instituto de Astrof\'\i sica de Andaluc\'\i a - CSIC Apdo. 3004 E-18080, Granada, Spain
\and 
IASF-INAF, Via Bassini 15, I-20133, Milano, Italy
\and
INAF-Osservatorio Astronomico di Bologna, via Ranzani 1, I-40127, Bologna, Italy
\and 
ESO, Karl-Schwarzschild-Str.2, D-85748 Garching b. M\"unchen, Germany
\and 
Oxford Physics, University of Oxford, Keble Road, Oxford, OX1\,3RH, UK
   }

\abstract{}{
This work aims to provide a first insight into the mass-metallicity ($MZ$) relation of star-forming galaxies at redshift  $z\sim 1.4$. 
To reach this goal, we present a first set of nine VVDS (VIMOS VLT Deep Survey) galaxies observed with the NIR integral-field 
spectrograph SINFONI on the VLT.
}{
Oxygen abundances are derived from empirical indicators based on the ratio between strong nebular emission-lines (H$\alpha$, [N\,\textsc{ii}]6584\AA{} and [S\,\textsc{ii}]6717,6731\AA{}\AA{}). Stellar masses are deduced from SED fitting with Charlot \& Bruzual (2003) population synthesis models, and star formation rates are derived from [O\,\textsc{ii}]3727\AA{} and H$\alpha$ emission-line luminosities.
}{
We find a typical shift of $0.2-0.4$~dex towards lower metallicities for the $z\sim1.4$ galaxies, compared to the $MZ$-relation in 
the local universe as derived from SDSS data. However, this small sample does not show any clear correlation 
between stellar mass and metallicity, unlike other larger samples at different redshift ($z\sim0$, $z\sim0.7$, and $z\sim2$).
Indeed, our galaxies lie just under the relation at $z\sim 2$, and show a small trend for more massive galaxies to be more metallic ($\sim 0.1$ logarithmic slope).
There are two possible explanations to account for these observations. 
First, our galaxies present higher specific star formation rates when compared to the global VVDS sample which could explain the particularly low metallicity of these galaxies as already shown in the SDSS sample. 
Second, inflow of metal-poor gas due to tidal interactions could also explain the low metallicity of these galaxies as three 
of these nine galaxies show clear signatures of merging in their velocity fields. Finally, we find that the metallicity of four galaxies is lower by $\sim 0.2$ to $0.4$~dex if we take into account the N/O abundance ratio in their metallicity estimate.
}{
}


\begin{document}

\maketitle

\section{Introduction}

The abundance of heavy elements in galaxies reflects the past history of star formation and the effects of gas 
exchanges (inflows and/or outflows) with the intergalactic medium. A characterization of the evolution of chemical 
abundances for galaxies of different masses is therefore essential to a complete model of galaxy formation 
that includes the physics of baryons \citep[see eg.][]{deLu04,deRo07a,FinDav08}.

First discovered for irregular galaxies \citep{leq79}, the mass-metallicity (hereafter $MZ$) relation has 
been investigated intensively \citep[Brodie \& Huchra 1981;][among others]{skill89, zar94, gar97}
and is now well established in the local universe ($z < 0.2$) thanks to the analysis of data from large spectroscopic surveys 
such as the Sloan Digital Sky Survey \citep[SDSS;][]{tre04,gall05,pan08} and the 
2 degree Field Galaxy Redshift Survey \citep[2dFGRS;][]{lam04}. The $MZ$ relation for local galaxies 
is steep for masses $\leq 10^{10.5}$ \msun\ and flattens at higher masses suggesting that the $MZ$ relation is mainly driven 
by the decrease of metal loss when stellar mass increases. These trends observed by \citet{tre04} 
and \citet{lam04} have been shown to extend to much lower galaxy masses \citep[$M < 10^9$ \msun; eg. Lee et al. 2006;][]{sav08}.
Recent studies focused on the dependence of the $MZ$ relation of SDSS galaxies with environment on small scales \citep[]{mdan08,ell08a} and on larger scales \citep{mouh07,coo08} and their star formation rate and galaxy sizes \citep{elli08b}.  

The evolution of the $MZ$ relation on cosmological timescales is now predicted by semi-analytical models of 
galaxy formation. Reliable observational estimates of the $MZ$ relation of galaxies at different epochs 
(and hence different redshifts) may thus provide important constraints on galaxy evolution scenarios. 
At increasing redshifts, as the strong rest-frame optical emission lines shift into the near-infrared (NIR) 
window, metallicities 
are typically based on smaller subsets of strong emission lines, through the use of empirically calibrated abundance 
indicators. Much progress has been made recently in assembling large samples of star-forming galaxies with abundance 
measurements at both intermediate redshifts \citep[$z < 1$;][]{sav05,maier05,liang06,
lam06,lam08,perez08a,cowBar08} 
and at $z > 2$ \citep{erb06,maio08,hay08,mann09}. 
However, chemical abundance measurements are only available for a limited number of galaxies at $z\sim 1-2$ \citep{shap05,maier06,liu08}. The rather unexplored $1 < z < 2$ redshift regime is one of particular importance 
in the history of the universe corresponding to i) the peak of star formation rate, and hence metal production, for the 
universe as a whole \citep[see eg.][]{lilly96,mad96,perGonz05,tresse07}, 
ii) the buildup of a significant fraction of the stellar mass in the universe \citep[eg.][]{drory05,arn07,poz07}, 
and iii) the emergence of the Hubble sequence of disks and spheroids.

We have undertaken an large observing  program at ESO-VLT aimed at probing the mass assembly and metallicity evolution 
of a representative sample of galaxies at $1 < z < 2$. The final goal of the 
MASSIV\footnote{\texttt{www.ast.obs-mip.fr/massiv/}} (Mass Assembly Survey with SINFONI in VVDS) project is to obtain 
a detailed description of the mix of dynamical types (rotating disks, spheroids, and mergers) at this epoch and to follow 
the evolution of fundamental scaling relations, such as the Tully-Fisher and $MZ$ ones, and therefore constrain 
galaxy evolution scenarios. In this paper we present the first results obtained by such a program focusing on the 
$MZ$ relation at $z \sim 1.4$.  A companion paper \citep{epi09}  is devoted to the kinematical analysis 
of galaxies at these redshifts using the same data. This work is based on a sample of nine galaxies 
selected in the VIMOS VLT Deep 
Survey \citep[VVDS;][]{lef05a} and observed in the NIR with the 3D-spectrograph SINFONI on the ESO/VLT \citep{eis03} during pilot observing runs. Throughout this paper we assume a standard $\Lambda$-CDM cosmology with $h = 0.7$, $\Omega_\Lambda = 0.7$, and 
$\Omega_m = 0.3$.

\section{Data description}

\subsection{Observations}
\label{sampleselect}
The selection of the sample and the SINFONI observations are fully described in \citet{epi09}. For the convenience of the reader, we summarize here the main steps.

We have used the VIMOS VLT Deep Survey to select galaxies with know spectroscopic redshifts in the range $1<z<2$. The VVDS is a complete magnitude selected sample avoiding the biases linked to more crude \emph{a priori} color selection techniques. It offers the advantage to combine a large sample with a robust selection function and secure spectroscopic redshifts necessary to engage into long single objects integrations with SINFONI being sure to observe the H$\alpha$ line outside of the OH sky emission lines.

The nine galaxies studied in this paper have been selected among those showing 
the strongest \oii\AA\ emission line (EW $> 50$ \AA\ and flux $> 5\times 10^{-17}$ ergs$^{-1}$cm$^{-2}$, as measured on VIMOS spectra) for H$\alpha$ to be easily detected in the near-IR.
These criteria for selecting late-type star-forming galaxies have been shown to be very 
efficient.
From the first observing runs, our success rate of selection has been around 85\%: 9 galaxies over 12 observed show strong rest-frame optical emission lines in SINFONI datacubes.
The other three galaxies have not been detected with SINFONI for the following reasons.
The redshift of one galaxy is such that H$\alpha$ is falling just in the gap between the $J$ and $H$ bands.
For another galaxy, the strength of [O\,\textsc{ii}]3727\AA\ was overestimated.
New measurements of this emission line show that this galaxy is now outside of the selection box.
The third non-detected galaxy had a wrong redshift.

Among the nine VVDS star-forming galaxies studied in this paper, five have been selected in the VVDS-22h wide field 
($17.5 \leq I_{AB} \leq 22.5$) and four in the VVDS-02h deep field ($17.5 \leq I_{AB} \leq 24.0$). Their basic 
properties (RA, DEC, redshift, $I$-band magnitude, etc) are listed in Table~\ref{carac}. These targets span a redshift range between $1.27\lesssim z\lesssim 1.53$. The NIR spectroscopic observations were acquired with the 3D-spectrograph SINFONI at ESO-VLT during two 4-nights run on September 5-8, 2005 (ESO run 75.A-0318) and November 12-15, 2006 (ESO run 78.A-0177). SINFONI was used in its seeing-limited mode, 
with the $0.25''\times 0.125''$ pixel scales leading to a field-of-view of 8\arcsec\ $\times$ 8\arcsec, and the H-band grism 
providing a spectral resolution $R \sim 4000$. Note also that two galaxies (VVDS220544103 and VVDS220584167) have also  been observed with the J-band grism. 
Conditions were not photometric and the mean seeing was around 0.65\arcsec. The total on-source integration times, as well as general information are listed in Table~\ref{carac}. For each Observation Block a hot telluric standard star has been observed. With this star we were able to flux-calibrate our observations with an accuracy of about 20\%. Details for data reduction are available in \citet{epi09}.



\begin{table*}
\begin{center}

\begin{threeparttable}
\caption{Properties of the nine galaxies observed with SINFONI.}\label{carac}

\begin{tabular}{ccccccc}
\hline\hline
Galaxy & RA(2000) & DEC(2000) & $t_\text{exp}$ (band)\tnote{1} & $I_{AB}$ & $z_\text{VIMOS}$\tnote{2} & $z_\text{SINFONI}$\tnote{3}  \\
& & & [hours] & & &  \\
\hline
VVDS020116027 & 02:25:51.2 & $-$04:45:07.5 & 1.7(H)       & $22.87$ & 1.5259 & 1.5302 \\
VVDS020182331 & 02:26:44.3 & $-$04:35:51.9 & 3(H)         & $22.73$ & 1.2286 & 1.2283 \\
VVDS020147106 & 02:26:45.5 & $-$04:40:46.9 & 2(H)         & $22.50$ & 1.5174 & 1.5195 \\
VVDS020261328 & 02:27:11.1 & $-$04:25:30.7 & 1(H)         & $23.90$ & 1.5291 & 1.5289 \\
VVDS220596913 & 22:14:29.2 & $+$00:22:15.7 & 1.75(H)      & $21.84$ & 1.2667 & 1.2662 \\
VVDS220584167 & 22:15:22.8 & $+$00:18:47.5 & 1.75(H),1(J) & $22.04$ & 1.4637 & 1.4659 \\
VVDS220544103 & 22:15:25.5 & $+$00:06:40.0 & 1(H),1(J)    & $22.47$ & 1.3970 & 1.3966 \\
VVDS220015726 & 22:15:42.5 & $+$00:29:04.9 & 2(H)         & $22.47$ & 1.3091 & 1.2930 \\
VVDS220014252 & 22:17:45.5 & $+$00:28:40.3 & 2(H)         & $22.10$ & 1.3097 & 1.3101 \\
\hline
\end{tabular}

\begin{tablenotes}
\item[1] {Exposure time corresponding to the effective time spent on the source in each band ($H$ or $J$).}
\item[2] {Redshift based on the VIMOS spectrum.}
\item[3] {Redshift estimate from the position of \ha{} in SINFONI spectra.}
\end{tablenotes}

\end{threeparttable}

\end{center}
\end{table*}

\subsection{Measurement of emission lines}

\begin{figure}
 \begin{center}
 \includegraphics[width=\linewidth]{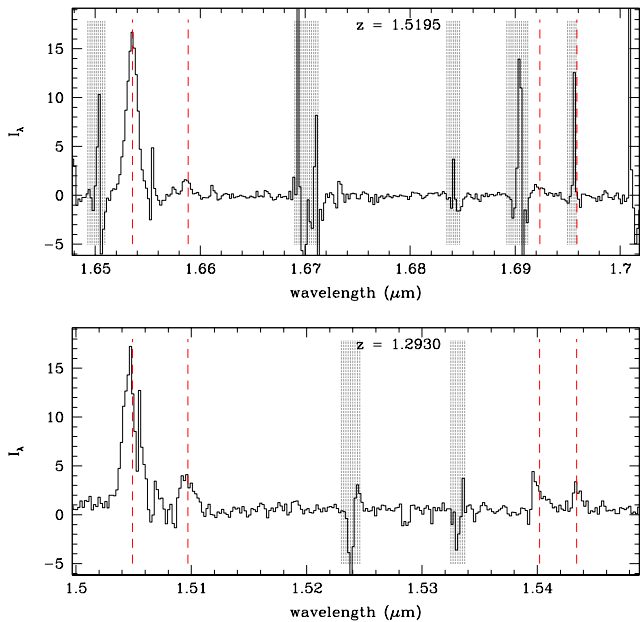}
\caption{Examples of 1D spectrum for two galaxies VVDS220015726 (bottom) and VVDS020147106 (top) observed with SINFONI in the $H$ band. The spectral region of interest is centered around the \ha{}, \niia{}\AA\ and [S\,\textsc{ii}]6717,6731\AA\ emission lines. The position of these lines is indicated by vertical dotted lines. The other lines with P-cygni profiles visible in the spectra are residuals of OH sky lines subtraction, they are shown within grey zones}\label{spectres}
 \end{center}
\end{figure}

To perform the spectrophotometric analysis of the reduced datacubes and measure the detected emission lines, we first produced 1D spectra from the 3D data. It has been performed in two steps: first, we produced a 2D pseudo-slit spectrum from the datacube (as if we placed a long-slit over the spatial extent of the object), and then we reduced the 2D-spectrum in a 1D-one, taking care of the  spatial extent of the galaxy and of the evolution of its position through the dispersion axis. The first step has been done with the \textsf{QFitsView} software\footnote{\texttt{http://www.mpe.mpg.de/$\sim$ott/QFitsView/}} which is a tool to visualize and analyse datacubes. The second step was performed using the \textsf{IRAF} software, specifically applying the \texttt{apall} task of the \texttt{noao>twodspec>apextract} package. Figure~\ref{spectres} shows examples of 1D spectra for the galaxies VVDS020147106 and VVDS220015726.

At the distance of the sample galaxies, the \ha{}, \niia{}\AA\ and [S\,\textsc{ii}]6717,6731\AA\ lines are redshifted into the SINFONI $H$ band. We were thus 
able to measure these lines in quasi-all our sample galaxies. For two galaxies we had observations in the $J$ band, 
which enables to detect and measure [O\,\textsc{iii}]4959,5007\AA\ and H$\beta$ lines for VVDS220544103. Unfortunately, the redshift of the second galaxy 
VVDS220584167 makes the reliable measurement of these lines impossible as they fall at the location of the bright OH-lines.
We measured the flux in the different lines fitting a gaussian to the 1D spectrum of each galaxies (within the \texttt{splot} task in \textsf{IRAF}). Fluxes and errors are obtained with the \textsf{deblending} mode using 100 simulations for the poissonian noise model (see the \texttt{splot} task help for more information). Results are shown in Table \ref{mesures}. Two reasons can account for the impossibility to measure a line: either an OH sky line falls right upon the line of interest, or the intensity of the line is too low to be detected. Over the nine galaxies, \niia{}\AA{} is not detected for two (VVDS020261328 and VVDS020182331).

\begin{table*}
\begin{center}

\begin{threeparttable}
\caption{Emission-line flux (in $10^{-17}$erg s${}^{-1}$ cm${}^{-2}$) for the nine galaxies observed with SINFONI.}\label{mesures}

\begin{tabular}{ccccccccc}
\hline\hline
Galaxy    & [O\,\textsc{ii}]3727\tnote{1} &  H$\beta$ & [O\,\textsc{iii}]4959 & \oiii & \ha  &  \niia & [S\,\textsc{ii}]6717 & [S\,\textsc{ii}]6731 \\
\hline
VVDS020116027 & $6.03\pm1.11$  &\--- & \--- & \--- & $6\pm1$ & $1\pm0.3$   & \--- & $0.7\pm 0.4$\\
VVDS020182331 & $19.85\pm1.41$ &\--- & \--- & \--- & $55\pm14$ & \--- & \--- & \--- \\
VVDS020147106 & $8.53\pm1.52$  &\--- & \--- & \--- & $22 \pm 6$ & $2.9\pm 1$ &  $1.4\pm 0.6$ & \--- \\
VVDS020261328 & $10.20\pm2.01$ &\--- & \--- & \--- & $5\pm1$ &\---   &\--- & \--- \\
VVDS220596913 & $18.55\pm1.78$ &\--- & \--- & \--- & $33\pm6$ & $4.7\pm1$ &   \--- & \---\\
VVDS220584167 & $17.90\pm1.83$ &\--- & \--- & \--- & $48\pm10$ & $11\pm3$ &  $6\pm 3$ & $7\pm3$\\
VVDS220544103 & $25.32\pm2.97$ &$38\pm9$ & $46\pm12$ & $114\pm14$ & $161\pm29$ & \---   & \--- & \---\\
VVDS220015726 & $9.47\pm1.41$  &\--- & \--- & \--- & $26\pm3$ & $4.6\pm 1$ &   \--- & \---\\
VVDS220014252 & $37.54\pm1.73$ &\--- & \--- & \--- & $212\pm27$ & $48\pm6$&  $15\pm2$ & \--- \\
\hline
\end{tabular}

\begin{tablenotes}
\item[1] as measured in the optical VIMOS spectra
\end{tablenotes}

\end{threeparttable}
\end{center}
\end{table*}

The measured values of $\log(\text{\niia}/\text{H}\alpha)$ range from $-1.3$ to $-0.64$, which indicates that the sources are star-forming galaxies with at most a low AGN contamination.We also checked for possible X-ray identification of AGN in our sample using the extensive HESEARC archive. Neither XMM nor Chandra source has been identified in the vincinity of our targets further excluding any AGN contamination in our sample. This issue is further discussed in \citet{epi09}.

\section{Stellar mass and metallicity estimates}

\subsection{Stellar mass}\label{mass}
The procedure to derive stellar mass of our galaxies is based on the comparison between observations, basically the photometric SED of galaxies, and population synthesis models. 
More precisely, $M^\star$ estimates were obtained with the GOSSIP spectral energy distribution (SED) modeling software \citep{franz08}.
For the SED fitting the multi-band photometric observations available in the VVDS fields were used, including $BVRI$ data from the CFHT, $UBVRZs$ data from the CFHT Legacy Survey, $J$- and $K$-bands data from SOFI at the NTT and from the UKIDSS survey, and the VVDS-Deep spectra.
The stellar population models used for the fit were generated with the BC03 population synthesis code (Bruzual \& Charlot 2003), assuming a set of ``delayed'' star formation histories \citep[see][for details]{gav02}, and a Salpeter IMF.
The values and erros bars of the stellar masses in Table~\ref{metali} are the median of the Probability Distribution Function and its confidence regions. More details can be found in \citet{epi09}.


\subsection{Metallicity}\label{metall}

To determine the gas-phase oxygen abundance of the nine VVDS galaxies, we made use of well-known empirical abundance indicators. Indeed, at these redshifts, we only have access to the brightest recombination and forbidden emission lines of the ionized interstellar gas, preventing any accurate metallicity estimate based on faint [O\,\textsc{iii}] auroral lines. Different calibrations or empirical indicators have been proposed over the past 20 years, each of them taking into account different set of emission lines. The most commonly used, but which has the drawback of being degenerated, is certainly $R_{\rm 23} = ([\text{O\,\textsc{ii}}]3727 + [\text{O\,\textsc{iii}}]4959,5007) / \text{H}\beta$, first introduced by Pagel et al. (1979), and re\--calibrated later in many articles \citep{Kewley02,Nagao06,Yin06}.

Considering that the nine galaxies have been observed with SINFONI in the $H$ band, we had access to H$\alpha$ and \niia{}\AA\ emission-line fluxes for six galaxies (we did not detect \niia{} for three galaxies). This led us to use the N2 indicator, defined as follows:
\begin{equation}
\text{N2} = \log\frac{\text{\niia}}{\text{H}\alpha}
\end{equation}
which is subject to many calibrations for metallicity estimates \citep{Kewley02,den02,Nagao06,Yin06,perez08b}. In our cases, the N2 parameter offered two main advantages to derive metallicity. Due to the proximity of the H$\alpha$ and \niia{}\AA\ lines, the N2 ratio is independant of reddening correction and uncertainty in relative flux calibration. Recently, \citet{perez08b} (hereafter PMC08) provided
a new calibration of this parameter based on a sample of objects with a direct determination of the corresponding ionic abundances:
\begin{equation}\label{calPerez}
12+\log O/H = 9.07 + 0.79\times\text{N2} 
\end{equation}
Nevertheless, this parameter presents a large scatter, which can be due to many processes, as explained in \citet{perez05}.
One possible explanation is the dependence of the calibration on the nitrogen-to-oxygen abundance ratio $N/O$. PMC08 investigated how the N2 abundance indicator can be altered by the value of the $N/O$ ratio.
They found the following deviation:
\begin{equation}\label{difference}
\Delta(O/H) = 0.5 \log N/O + 0.66
\end{equation}
Following their method, we had to determine the nitrogen-to-oxygen abundance ratio. \citet{perez08b} propose an empirical calibration for it, derived from the same sample of H\,\textsc{ii} regions they used to derive equations (\ref{calPerez}) and (\ref{difference}). For our purpose, considering the emission lines we have in our spectra, we have used use the $\log\dfrac{\text{\niia}}{\text{[S\,\textsc{ii}]6717,6731}} = \mathrm{N2S2}$ ratio leading to the following calibration:
\begin{equation}\label{nsuro}
 \log(N/O) = 1.26\times\mathrm{N2S2} - 0.86
\end{equation}
In this work, we first calculated the metallicities using equation~(\ref{calPerez}): the results are listed in Table \ref{metali}. Unfortunately, the measurement of the sulfur lines ([S\,\textsc{ii}]6717,6731\AA{}) is very difficult, and we sometimes only had a measurement for one of the two lines, so we derived the other using the relation (\ref{sulfur}) which is valid if we assume a typical density of 100~cm${}^{-2}$ \citep{ram93}:
\begin{equation}\label{sulfur}
\frac{[\text{S}\,\textsc{ii}]6717}{[\text{S}\,\textsc{ii}]6731}=1.32
\end{equation}
In \citet{oster}, Fig 5.3 p.134, we can see that at these typical densities, the value of the ratio changes by 0.1 at most. We were thus able to ``correct'' the metallicities of 4 galaxies, the ones for which we had a measurement of [S\,\textsc{ii}]6717,6731\AA{}. The derived metallicities, corrected for the $N/O$ ratio, are also listed in Table \ref{metali}.

For three galaxies (VVDS220544103, VVDS020261328 and VVDS020182331) we did not measure the [N\,\textsc{ii}]6584\AA{} line. However, observations 
in the $J$ band were available for the first galaxy (VVDS220544103) enabling to measure [O\,\textsc{iii}]4959,5009\AA{} and 
H$\beta$ emission lines (see Table~\ref{mesures}). With H$\alpha$, measured in the $H$ band, and [O\,\textsc{ii}]3727\AA{} 
(from the VIMOS spectrum) we had thus in hand five emission lines for this galaxy to derive metallicity using the $R_{\rm 23}$ 
indicator. The ratio of the Balmer lines $\text{H}\alpha/\text{H}\beta$ gives an estimate of the attenuation by the dust. 
Assuming a classical value for the intrinsic ratio $F(\text{H}\alpha)/F(\text{H}\beta)$ of 2.85, our measurements of the 
H$\alpha$ and H$\beta$ lines gave an extinction coefficient of $C=0.31\pm0.33$. We were then able to correct the emission-line fluxes 
for reddening using the Calzetti extinction law \citep{cal01} and use these values to derive metallicity. We obtained 
$\log R_{\rm 23}=0.702$ which led to an estimate of the metallicity of $12+\log(O/H)=8.78$, using the calibration 
of \citet{zar94}. \citet{kd08} provide a way to translate this value to T04 (the calibration used in \citet{tre04}). 
We finally obtained $12+\log(O/H)|_\text{T04}=8.74$ for VVDS220544103.

\begin{table*}
\begin{center}
\begin{threeparttable}

\caption{Metallicity ($12+\log(O/H)$) and stellar mass estimates for the 7 galaxies observed with SINFONI.}\label{metali}

\begin{tabular}{cccccccc}
\hline\hline
Galaxy  & $Z_\text{PMC08}$\tnote{1} & $Z_\text{PMC08 cor.}$\tnote{2} & $Z_{\text{PMC08}\rightarrow\text{T04}}$\tnote{3}  &  $Z_{\text{PMC08 cor.}\rightarrow\text{T04}}$\tnote{3} & $\log(M^\star/M_\odot)$\tnote{4} & $\log(N/O)$\tnote{5} & $\log(\text{SSFR yr}^{-1})$\tnote{6} \\
\hline
VVDS020116027 & $8.46\pm0.09$ &  $8.43\pm0.27$ & $8.66\pm0.12$ & $8.62\pm0.37$  & $10.09^{+0.38}_{-0.21}$ & $-1.13\pm0.49$ & $-8.72\pm0.20$ \\
VVDS020147106 & $8.37\pm0.16$ &  $8.15\pm0.16$ & $8.56\pm0.2$  & $8.29\pm0.19$  & $10.23^{+0.01}_{-0.18}$ & $-0.77\pm0.22$ & $-8.22\pm0.31$ \\
VVDS220596913 & $8.40\pm0.11$ &  \---          & $8.59\pm0.14$ & \---           & $10.93^{+0.12}_{-0.29}$ & \---           & $-8.94\pm0.21$ \\
VVDS220584167 & $8.56\pm0.13$ &  $8.44\pm0.24$ & $8.80\pm0.18$ & $8.64\pm0.33$  & $11.08^{+0.28}_{-0.23}$ & $-0.95\pm0.43$ & $-8.73\pm0.24$ \\
VVDS220544103 & \---          &  \---          & \---          & \---           & $10.71^{+0.16}_{-0.41}$ & \---           & $-7.81\pm0.21$ \\
VVDS220015726 & $8.44\pm0.11$ &  \---          & $8.63\pm0.13$ & \---           & $10.79^{+0.29}_{-0.13}$ & \---           & $-8.88\pm0.13$ \\
VVDS220014252 & $8.56\pm0.08$ &  $8.20\pm0.12$ & $8.80\pm0.10$ & $8.35\pm0.15$  & $10.78^{+0.07}_{-0.39}$ & $-0.53\pm0.22$ & $-7.83\pm0.30$ \\
\hline
\end{tabular}

\begin{tablenotes}
\item[1] metallicity estimate using PMC08 indicator.
\item[2] metallicity estimate using PMC08 indicator and $N/O$ correction with the sulfur lines.
\item[3] corresponding metallicities converted into T04 metallicity estimator.
\item[4] stellar mass in solar units (see \S~\ref{mass}).
\item[5] $\log(N/O)$ estimated with the N2S2 ratio.
\item[6] specific SFR in yr$^{-1}$, derived with $L(\text{H}\alpha)$.
\end{tablenotes}

\end{threeparttable}
\end{center}
\end{table*}

\section{The Mass-Metallicity relation}

\subsection{The $MZ$ relation}\label{mzr}
\begin{figure}
\centering
\includegraphics[width=\linewidth]{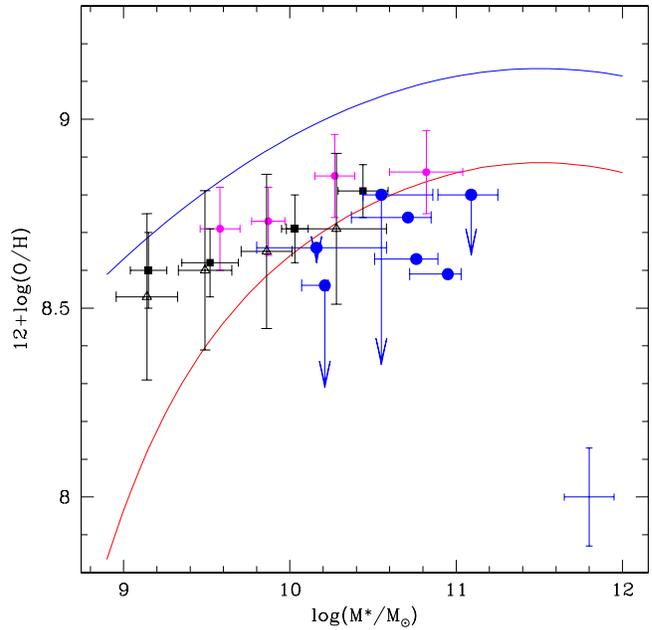}
\caption{Mass \--- metallicity relation for the sample of seven VVDS galaxies (blue circles) at $z\sim 1.4$. The arrows show how the 
metallicities change when we include the correction of $\log(N/O)$, following \citet{perez08b} prescription. The error bars at the bottom right corner show typical uncertainties of our data.
The blue curve is the $MZ$ relation in the local universe as derived from the SDSS by \citet{tre04}. The red curve is the $MZ$ relation at $z\sim2$ by \citet{erb06}. The $MZ$ relation at $z\sim0.7$ \citep{lam08} and $z\sim 1$ \citep{perez08a} from the VVDS sample, and at $z\sim 1$ from zCOSMOS (Contini et al., \emph{in prep.}) are shown in black open triangles, black squares, and magenta circles respectively. All metallicities have been translated into T04 to facilitate comparisons.}\label{mz}
\end{figure}

Figure~\ref{mz} shows the position of the seven galaxies at $z\sim1.4$ in the $MZ$-plane (blue circles) using the metallicities and the stellar masses listed in Table~\ref{metali}. In order to compare with the $MZ$ relations previously derived at different redshifts, we also draw in this figure the $MZ$ relations obtained in the local Universe \citep{tre04} as derived from the the SDSS, at $z\sim0.7$ \citep{lam08} and $z\sim 1$ \citep{perez08a} from the VVDS sample, at $z\sim 1$ from zCOSMOS (Contini et al., \emph{in prep}), and at $z\sim 2$ \citep{erb06}. However, such a comparison requires great precautions, as pointed out in \citet{kd08}. In their study, \citet{kd08} investigate the influence of different metallicity calibrations on the $MZ$ relation and showed that the absolute value of the metallicity can vary up to 0.8 dex depending on the calibration used. Taking this remark into consideration, our purpose being to compare the $MZ$ relation at $z\sim1.4$ to the ones at others epochs, we had to be consistent with the calibrations used in other studies.

Firstly, at $z\sim 0$ \citet{tre04} used an indicator based on the Bayesian calibration CL01 \citep[for][]{cl01} \--- we will refer 
as T04 \--- which make use of a large set of emission lines to derive metallicity. In order to translate our metallicities into 
the T04 calibrator, we first converted to \citet{den02} N2 calibration and then applied the transformation given in \citet{kd08} to get 
T04 metallicities. If $x$ stands for the metallicity estimate of \citet{perez08b} and $y$ for that of T04, we then have:
\begin{equation}
 y = -0.217469 x^3 + 5.80493 x^2 - 50.2318 x + 149.836
\end{equation}
The resulting points in Fig.~\ref{mz} are plotted as blue circles.

Secondly, at $z\sim 2$ \citet{erb06} used another calibration of the N2 indicator: PP04 \citep{pet04}. To be consistent we chose to convert their relation into T04. In \citet{kd08} we found the way to convert from PP04 to T04, as follows (if $x$ stands for the estimate of metallicity with PP04 and $y$ with T04):
\begin{equation}
y = 2.6766690 x^3 - 68.471750 x^2 + 585.1750 x - 1661.9380
\end{equation}
The resulting curve is shown in Fig.~\ref{mz} in red. The three other relations \citep[][Contini et al. \emph{in prep.}]{lam08,perez08a} are already calibrated with T04 estimator.

While \citet{tre04}, \citet{erb06}, \citet{lam08} and \citet{perez08a} found a monotonic relation between the metallicity and stellar mass \--- more massive galaxies have a higher nebular metallicity than less massive ones \--- our small sample of galaxies at $z\sim 1.4$ does not show any obvious correlation between metallicity and stellar mass, even though one could imagine $Z$ to increase with stellar mass.
Actually, if we fit a line through our 7 points, we obtain a positive slope ($12+\log(O/H) = 0.101\log(M^\star/M_\odot)+7.61$).
The mean metallicity for this sample of seven galaxies is $12+\log(O/H)=8.62$ with a scatter of $0.22$~dex. 
Note however that the $z\sim 1.4$ galaxies are lying below the $MZ$ relation derived in the local universe from SDSS data (blue line in Fig.~\ref{mz}).
The shift in metallicity is roughly $-0.3$~dex, which clearly confirms the evolutionary trend of the galaxies metallicity with cosmic time for a given stellar mass.
Six out of seven galaxies have metallicities lower than the median $z\sim 2$ relation of \citet{erb06} which seems to draw an upper limit in metallicity for our sample.
We can wonder why our galaxies seem poorly metallic, and in particular less metallic than \citet{erb06} relation.
 
\begin{figure}
 \begin{center}
 \includegraphics[width=\linewidth]{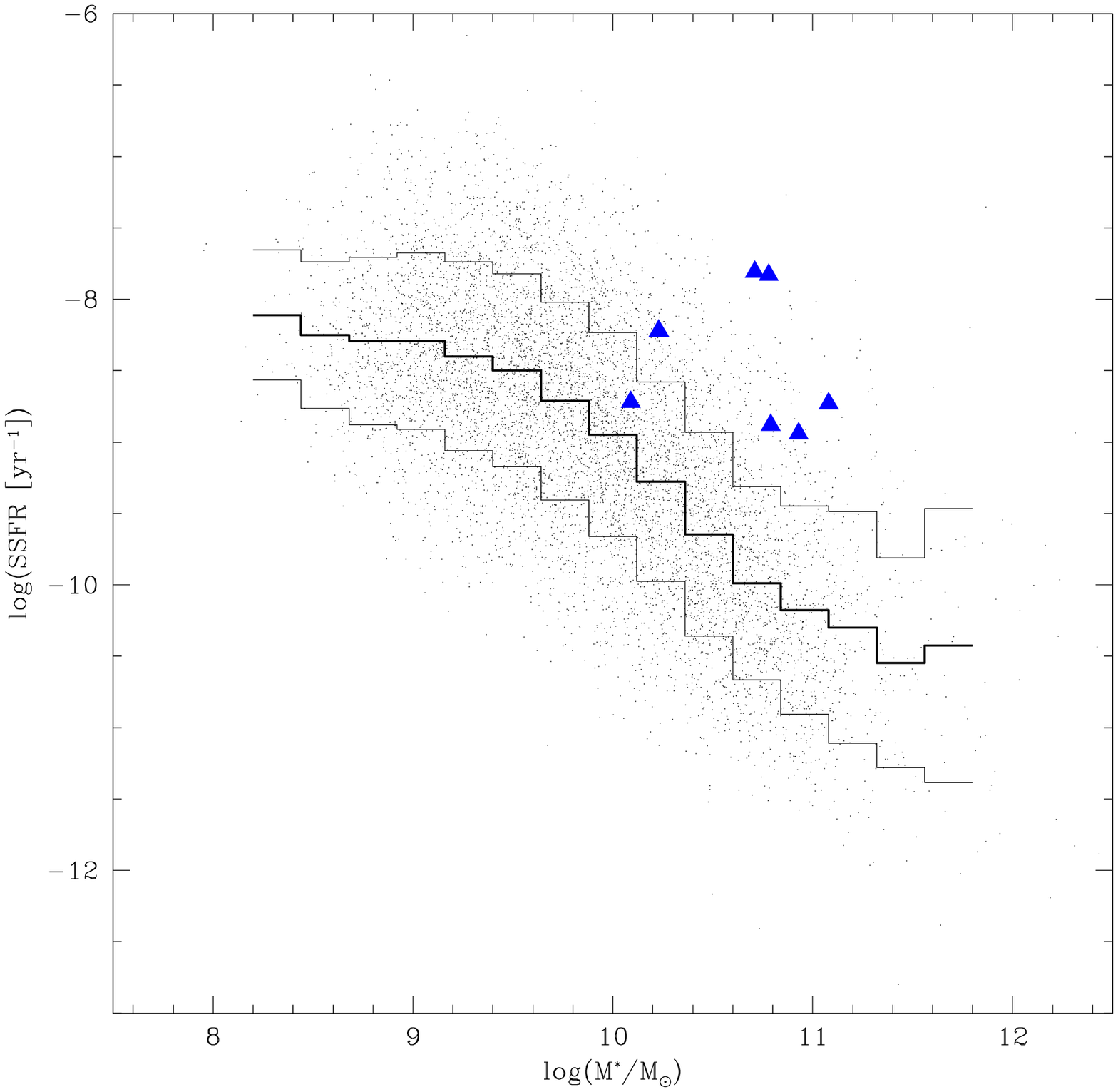}
\caption{Specific star formation rate vs. stellar mass for VVDS galaxies in the 02h and 22h fields (see \S~\ref{mass} for mass estimates).
Values are obtained with SFR derived from [O\,\textsc{ii}]3727\AA{} luminosity, with the calibration of \citet{arg09}.
The histograms represent the mean specific SFR distribution (bold black line) and 1$\sigma$ scatter (thick black lines).
The blue triangles stand for the seven galaxies of our sample, which SFR have been derived from H$\alpha$ luminosity, as in \citet{arg09}.}\label{ssfr}
\end{center}
\end{figure}

Recently \citet{elli08b} investigated the systematic effects of specific star formation rate (SSFR) on the $MZ$ relation. 
At a given mass, SDSS galaxies with high SSFR have in average lower metallicities than galaxies with low SSFR. 
We thus derived SSFR for the 7 galaxies in our sample, using the H$\alpha$ luminosity, the SFR calibration by \citet{arg09} [eq. 17]  and the stellar mass given by the SED fitting.
To be able to compare with the entire VVDS sample, we also derived specific SFR for galaxies in the 22h~$+$~02h ``wide'' fields, for which we had a measure of $F(\mathrm{[O\,\textsc{ii}]3727\AA{}})$ and an estimate of the stellar mass. 
\citet{arg09} provide a calibration of the SFR based on the [O\,\textsc{ii}]3727\AA{} line [eq. 18] which is consistent with the H$\alpha$ calibration. 
In Fig. \ref{ssfr} we show the SSFR as a function of stellar mass for the full VVDS 22h~$+$~02h ``wide'' sample, together with our $z\sim1.4$ VVDS galaxies. 
As one can see our galaxies lie on top of the mean distribution in SSFR, even out of the 1$\sigma$ limits (except for one), which could explain why their metallicity is so low. As a matter of fact our sample is clearly biased toward high specific star formation rates.
 Several explanations are proposed for this trend \citep[see][for more details]{elli08b}: galactic winds could be more efficient in extracting metals at higher SSFRs, inflows of metal-poor gas could increase the SFR, etc\dots{}
The fact that these seven galaxies show on average higher SSFRs than the global VVDS sample is not surprising as they were selected on the basis of the intensity of the [O\,\textsc{ii}]3727\AA{} line (see Sect.~\ref{sampleselect}).
This \emph{bias} could explain why the $MZ$ relation we find at $z\sim1.4$ seems shifted toward low metallicities.
Another explanation, which could be linked to the SSFR one, has been recently proposed by \citet{pee09}.
These authors claim that the presence of tidal interactions in majors mergers could possibly account for a fading in oxygen abundance as metal-poor gas may inflow from large radii.
In our sample, VVDS220596913, VVDS220544103 and VDS020116027 seem indeed to undergo tidal interactions as revealed by their peculiar velocity fields \citep[see][]{epi09}.

\subsection{Influence of the $N/O$ ratio}
\begin{figure}
\centering
\includegraphics[width=\linewidth]{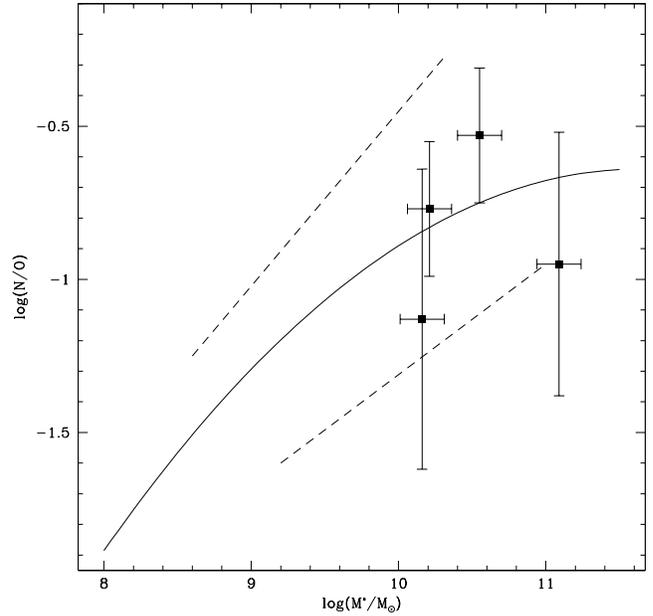}
\caption{Relation between stellar mass and the $N/O$ abundance ratio. The solid line is the median relation of the SDSS star-forming galaxies in the local universe\citep[see][fig. 14]{perez08b}.The dashed lines indicate the scatter of this relation. The filled black squares correspond to the four VVDS galaxies for which we have measurements of the sulfur lines to derive $N/O$ ratio.}\label{novsm}
\end{figure}

Figure \ref{novsm} shows the relation between the stellar mass and the nitrogen-to-oxygen abundance ratio $N/O$ for the SDSS star-forming galaxies in the local universe \citep[][their Fig. 13]{perez08b}.
The four VVDS galaxies at $z\sim 1.4$ for which we were able to derive $\log(N/O)$ are also shown in this figure as filled squares.
Even if the dispersion around the relation defined by the SDSS sample is high, it is clear that the four VVDS galaxies are consistent with this relation. 
The influence of the nitrogen-to-oxygen abundance ratio on the metallicity is shown on Fig.~\ref{mz} with the downwarding arrows. 
As noticed by \citet{perez08b} the metallicities are significantly shifted to lower values for high $N/O$ abundances ratios.
This is especially the case for VVDS020147106 and VVDS220014252 which exhibit the highest $N/O$ values. 
Looking at Fig.~\ref{mz} we must keep in mind the important error bars going up to nearly $0.5$~dex for the $N/O$ ratio (see Table~\ref{metali}).
Nevertheless, even if we take into account these errors, the downwarding shifts in metallicity \--- for VVDS020147106 and  VVDS220014252 \--- are too important to deny any effective decline in oxygen abundance: indeed the drops in oxygen abundance are larger than the error bars themselves, which guarantee an effective decline. 

\section{Conclusion and prospects}
We have determined the stellar mass and metallicity of seven VVDS star-forming galaxies at $z\sim 1.4$ observed in the NIR with the integral-field spectrograph SINFONI.
Keeping in mind the uncertainties inherent to the metallicity and mass estimates, and considering also that our small sample does not allow to draw any firm conclusion, we can however draw the following points.
Our sample of seven $z\sim 1.4$ galaxies shows if any, a mild correlation between metallicity and stellar mass ($\sim 0.1$ logarithmic slope).
The mean metallicity is $12+\log(O/H)=8.62$ with a scatter of $0.22$~dex.
All the galaxies in our sample are just under the relation of \citet{erb06} at $z\sim 2$, except for one galaxy that lies just above.
The low metallicity measured in these galaxies at $z\sim 1.4$ could be due to their high specific star formation rate and/or to the tidal interactions allowing for inflow of metal-poor gas in the cases of VVDS220596913, VVDS220544103 and VDS020116027.
Accounting for the dependence of oxygen abundance estimate on the $N/O$ abundance ratio, as suggested by \citet{perez08b}, would lower significantly the metallicity for two galaxies showing the largest $N/O$ abundance ratios.
When completed, the MASSIV survey will gather a representative sample of roughly 80 galaxies observed with SINFONI within $1<z<2$.
It will allow to shed new light on these issues enabling to probe the dynamical and physical properties of these galaxies.
Concerning the $MZ$ relation, a statistical study will be undertaken over a large range of stellar masses, computing median metallicities in bins of mass and deriving the median relation between  $12+\log(O/H)$ and $M^\star$ and its intrinsic scatter.
Comparing these observations to the predictions of galaxy evolution models \citep[\emph{e.g.}][]{deRo07a} will also be an major issue to investigate within MASSIV.

\begin{acknowledgements}
We wish to thank the ESO staff at Paranal Observatory and especially the SINFONI team at VLT for their support during observations.
We also thank the referee for the constructive comments which help to improve the quality of this paper.
This work has been partially supported by the CNRS-INSU Programme National de Galaxies and by the french ANR grant ANR-07-JCJC-0009.
\end{acknowledgements}

\bibliographystyle{aa}
\bibliography{11994}

\end{document}